\documentclass[12pt]{article}
\usepackage{graphicx}
\usepackage{lscape}
\usepackage{citesort}
\usepackage{amssymb}
\usepackage{appendix}
\usepackage{multirow}

\begin{document}
\def\qq{\langle \bar q q \rangle}
\def\uu{\langle \bar u u \rangle}
\def\dd{\langle \bar d d \rangle}
\def\sp{\langle \bar s s \rangle}
\def\GG{\langle g_s^2 G^2 \rangle}
\def\Tr{\mbox{Tr}}
\def\figt#1#2#3{
        \begin{figure}
        $\left. \right.$
        \vspace*{-2cm}
        \begin{center}
        \includegraphics[width=10cm]{#1}
        \end{center}
        \vspace*{-0.2cm}
        \caption{#3}
        \label{#2}
        \end{figure}
    }

\def\figb#1#2#3{
        \begin{figure}
        $\left. \right.$
        \vspace*{-1cm}
        \begin{center}
        \includegraphics[width=10cm]{#1}
        \end{center}
        \vspace*{-0.2cm}
        \caption{#3}
        \label{#2}
        \end{figure}
                }

\def\ds{\displaystyle}
\def\beq{\begin{equation}}
\def\eeq{\end{equation}}
\def\bea{\begin{eqnarray}}
\def\eea{\end{eqnarray}}
\def\beeq{\begin{eqnarray}}
\def\eeeq{\end{eqnarray}}
\def\ve{\vert}
\def\vel{\left|}
\def\ver{\right|}
\def\nnb{\nonumber}
\def\ga{\left(}
\def\dr{\right)}
\def\aga{\left\{}
\def\adr{\right\}}
\def\lla{\left<}
\def\rra{\right>}
\def\rar{\rightarrow}
\def\lrar{\leftrightarrow}
\def\nnb{\nonumber}
\def\la{\langle}
\def\ra{\rangle}
\def\ba{\begin{array}}
\def\ea{\end{array}}
\def\tr{\mbox{Tr}}
\def\ssp{{\Sigma^{*+}}}
\def\sso{{\Sigma^{*0}}}
\def\ssm{{\Sigma^{*-}}}
\def\xis0{{\Xi^{*0}}}
\def\xism{{\Xi^{*-}}}
\def\qs{\la \bar s s \ra}
\def\qu{\la \bar u u \ra}
\def\qd{\la \bar d d \ra}
\def\qq{\la \bar q q \ra}
\def\gGgG{\la g^2 G^2 \ra}
\def\q{\gamma_5 \not\!q}
\def\x{\gamma_5 \not\!x}
\def\g5{\gamma_5}
\def\sb{S_Q^{cf}}
\def\sd{S_d^{be}}
\def\su{S_u^{ad}}
\def\sbp{{S}_Q^{'cf}}
\def\sdp{{S}_d^{'be}}
\def\sup{{S}_u^{'ad}}
\def\ssp{{S}_s^{'??}}

\def\sig{\sigma_{\mu \nu} \gamma_5 p^\mu q^\nu}
\def\fo{f_0(\frac{s_0}{M^2})}
\def\ffi{f_1(\frac{s_0}{M^2})}
\def\fii{f_2(\frac{s_0}{M^2})}
\def\O{{\cal O}}
\def\sl{{\Sigma^0 \Lambda}}
\def\es{\!\!\! &=& \!\!\!}
\def\ap{\!\!\! &\approx& \!\!\!}
\def\ar{&+& \!\!\!}
\def\ek{&-& \!\!\!}
\def\kek{\!\!\!&-& \!\!\!}
\def\cp{&\times& \!\!\!}
\def\se{\!\!\! &\simeq& \!\!\!}
\def\eqv{&\equiv& \!\!\!}
\def\kpm{&\pm& \!\!\!}
\def\kmp{&\mp& \!\!\!}
\def\mcdot{\!\cdot\!}
\def\erar{&\rightarrow&}

% .........................................................

\def\simlt{\stackrel{<}{{}_\sim}}
\def\simgt{\stackrel{>}{{}_\sim}}

% .........................................................

\title{
         {\Large
                 {\bf
Properties of triply heavy spin--3/2 baryons
                 }
         }
      }

\author{\vspace{1cm}\\
{\small T. M. Aliev \thanks {e-mail:
taliev@metu.edu.tr}~\footnote{permanent address:Institute of
Physics,Baku,Azerbaijan}\,\,, K. Azizi \thanks {e-mail:
kazizi@dogus.edu.tr}\,\,, M. Savc{\i} \thanks
{e-mail: savci@metu.edu.tr}} \\
{\small Physics Department, Middle East Technical University,
06531 Ankara, Turkey }\\
{\small$^\ddag$ Physics Department,  Faculty of Arts and Sciences,
Do\u gu\c s University,} \\
{\small Ac{\i}badem-Kad{\i}k\"oy,  34722 Istanbul, Turkey}}

\date{}

\begin{titlepage}
\maketitle
\thispagestyle{empty}

\begin{abstract}

The masses and residues of the triply heavy spin--3/2 baryons are
calculated in framework of the QCD sum rule approach.
The obtained results are compared with the existing theoretical
predictions in the literature.
\end{abstract}
~~~PACS number(s): 11.55.Hx,  14.20.-c, 14.20.Mr, 14.20.Lq
\end{titlepage}

\section{Introduction}

The quark model predicts heavy baryons containing  single, doubly
or triply heavy charm or bottom quarks having either spin--1/2 or
spin--3/2. So far, all heavy baryons with single charm quark have
been discovered in the experiments. Some of the heavy baryons with
single bottom quark like $\Lambda_b$, $\Sigma_b$, $\Xi_b$ and
$\Omega_b$ baryons with spin--1/2 as well as $\Sigma_b^\ast$
baryon with spin--3/2 have been observed in the experiments (for a
review see for instance \cite{R01}). In 2012, the CMS
Collaboration reported the observation of the
$\Xi_b^\ast$ state with spin--3/2 \cite{R02}. 
At present, among all possible doubly heavy baryon states
only the spin--1/2  $\Xi_{cc}^+$
charmed baryon have experimentally been observed by the SELEX
Collaboration \cite{R03,R04,R05}. The experimental
attempts, especially at LHCb,
have still been continuing to complete the remaining members
of the heavy baryons with one, two or three heavy quarks predicted by
the quark model.

From the theoretical side, there are a lot of works in the literature
devoted to the spectroscopy and decay properties of the heavy baryons
containing single heavy quark.
There are also dozens of works dedicated to the study of the
properties of the doubly heavy baryons.
However, there are limited numbers of works devoted to the investigation
of the properties of the triply
heavy baryons. The masses of the triply heavy baryons have been studied in  
framework of the various approaches such as
effective field theory, lattice QCD,  bag model, various
quark models, variational approach, hyper central model, potential
model and Regge trajectory ansatz
\cite{R06,R07,R08,R09,R10,R11,R12,R13,R14,R15,R16,R17,R18,R19}.
The masses and residues  of the triply heavy spin--1/2 baryons for
the Ioffe current, as well as the masses of the triply heavy
spin--3/2 baryons,
are also calculated in \cite{R20,R21} in framework of the QCD sum rule approach.
In the present work, we extend our previous work on the spectroscopy of the
triply heavy spin--1/2 baryons for the general form of the interpolating current
\cite{R22} to
calculate the masses and residues of both positive and negative parity
triply heavy spin--3/2 baryons
in framework of the QCD sum rules. We compare our results on the
masses and residues of these baryons with the predictions of the existing
approaches in literature \cite{R08,R09,R10,R11,R12,R13,R14,R15,R20,R21}.
Information on the masses of the triply heavy baryons
can play essential role in understanding
the heavy quark dynamics.

The paper  is organized as follows. In the following section,
we derive  QCD sum rules for the masses
and residues of both negative and positive parity  triply heavy
spin--3/2 baryons. 
Section 3 is devoted to the analysis of the sum rules
for the masses and residues of the triple heavy baryons. This section
contains also a comparison
of the obtained results with the predictions of other approaches existing
in literature.

\section{QCD sum rules for the masses and residues of the triply
heavy spin--3/2 baryons}

In order to calculate the masses and residues of the triply heavy  spin--3/2 baryons
we start by the  following  two--point correlation function as the main
object of the method:
\bea
\label{edhbtt01}
\Pi_{\mu\nu} (q) = i \int d^4x e^{iqx} \lla 0 \vel {\cal T} \{ \eta_\mu (x)
\bar{\eta}_\nu (0) \} \ver 0 \rra ~,
\eea
where ${\cal T}$ is the time ordering operator,  $q$ is the four--momentum
of the corresponding triply heavy baryon and $\eta_\mu$
stands for its interpolating current, whose general form can written as
\bea
\label{edhbtt02}
\eta_\mu = {1\over \sqrt{3}} \epsilon^{abc} \Big\{2
(Q^{aT} C \gamma_\mu Q^{\prime b}) Q^c +
(Q^{aT} C \gamma_\mu Q^b) Q^{\prime c} \Big\}~,
\eea
where   $Q$ and $Q^\prime$
are   heavy quarks. The quark contents for all members of the triply
heavy spin--3/2 baryons are given in Table 1.
\begin{table}[htb]
\begin{center}
\begin{tabular}{|c|c|c|}\hline\hline
             Baryon        &$Q$ &  $Q'$    \\ \hline
      $\Omega^*_{bbc}$ &  $b$           &$c$     \\ \hline
      $\Omega^*_{ccb}$ &  $c$           &$b$     \\ \hline
 $\Omega^*_{bbb}$ &  $b$           &$b$     \\ \hline
 $\Omega^*_{ccc}$ &  $c$           &$c$     \\ \hline
        \end{tabular}
\end{center}
\caption{The quark contents of the  triply heavy spin--3/2 baryons.}
\end{table}

Having constructed the correlation function, our next task is construction
of the sum rule for the masses of the triply heavy baryons. In order to
construct the sum rules, this correlation function should be calculated in two
different ways:
in terms of hadronic
parameters (physical side), and in terms of QCD degrees of freedom
(QCD side).
Equating these two representations of the correlation function gives us the
sum rules for the masses of the triply heavy baryons
in terms of quark and gluon degrees of freedom.

Before calculating the correlation function
from the physical side,
we should mention that the interpolating current $\eta_\mu$ of the
triply  heavy baryons can interact not only with the positive and negative
parity spin--3/2
baryons, but also it couples to the triply heavy  spin--1/2 baryons with
both parities. In order to obtain reliable results we should eliminate the unwanted
spin--1/2 baryons' contributions.

Let us discuss the elimination of the contributions coming from
spin--1/2 states.
Using the parity and Lorentz covariance considerations,
the matrix elements of the interpolating current $\eta_\mu$ 
for the masses of the spin--3/2 triply heavy baryons
between the vacuum and the baryonic states, are defined as:
\bea
\label{edhbtt03}
\lla 0
\vel \eta_\mu \ver B_{(3/2)^+}(q) \rra \es \lambda_{(3/2)^+}
u_\mu (q)~, \nnb \\
\lla 0 \vel \eta_\mu \ver B_{(3/2)^-}(q) \rra \es \lambda_{(3/2)^-}
\gamma_5 u_\mu (q)~, \nnb \\
\lla 0 \vel \eta_\mu \ver B_{(1/2)^+}(q) \rra \es \lambda_{(1/2)^+}
\Bigg( {4 q_\mu \over m_{B_{(1/2)^+}}} + \gamma_\mu \Bigg) \gamma_5 u(q), \nnb \\
\lla 0 \vel \eta_\mu \ver B_{(1/2)^-}(q) \rra \es
\lambda_{(1/2)^-} \Bigg( {- 4 q_\mu \over m_{B_{(1/2)^-}}} +
\gamma_\mu \Bigg) u(q)~,
\eea
where $u(q)$ and $u_\mu(q)$ are the
Dirac and  Rarita--Schwinger spinors  for the
spin--1/2 and spin--3/2 baryons, respectively, and $\lambda_i$ are the
corresponding residues. Obviously, we see from Eq. (\ref{edhbtt03}) that 
the contributions coming from the
spin--1/2 states are proportional to $q_\mu$ or $\gamma_\mu$.
The physical part of the correlation function can be
calculated by saturating the correlation function with the ground state
baryons as follows:
\bea 
\label{edhbtt04}
\Pi_{\mu\nu} = {\lla 0
\vel \eta_\mu \ver B(q) \rra \lla B(q) \vel \bar{\eta}_\nu \ver 0
\rra\over q^2-m_B^2} + \cdots~,
\eea
where dots represent the
contributions coming from the higher states and continuum.
Using Eqs. (\ref{edhbtt03}) and (\ref{edhbtt04}), for the
physical part of the correlation function we get
\bea
\label{edhbtt06}
\Pi_{\mu\nu}(q) \es
{\lambda^2_{(3/2)^+} \over m_{(3/2)^+}^2 - q^2} (\rlap/{q} + m_{(3/2)^+})
\Bigg( g_{\mu\nu} - {1\over 3} \gamma_\mu \gamma_\nu -
{2 q_\mu q_\nu \over m_{(3/2)^+}^2} + {q_\mu\gamma_\nu - q_\nu\gamma_\mu
\over 3 m_{(3/2)^+}} \Bigg)~, \nnb \\
\ek {\lambda^2_{(3/2)^-} \over m_{(3/2)^-}^2 - q^2} \gamma_5 (\rlap/{q} + m_{(3/2)^-})
\Bigg( g_{\mu\nu} - {1\over 3} \gamma_\mu \gamma_\nu -
{2 q_\mu q_\nu \over m_{(3/2)^-}^2} + {q_\mu\gamma_\nu - q_\nu\gamma_\mu
\over 3 m_{(3/2)^-}} \Bigg) \gamma_5~, \nnb \\
\ek {\lambda^2_{(1/2)^+} \over m_{(1/2)^+}^2 - q^2} \Bigg( {4 q_\mu \over
m_{(1/2)^+}} +\gamma_\mu \Bigg) \gamma_5 (\rlap/{q} + m_{(1/2)^+})
\Bigg( {4 q_\nu \over m_{(1/2)^+}} +\gamma_\nu \Bigg) \gamma_5 \nnb \\
\ar {\lambda^2_{(1/2)^-} \over m_{(1/2)^-}^2 - q^2} \Bigg( - {4 q_\mu \over
m_{(1/2)^-}} +\gamma_\mu \Bigg) (\rlap/{q} + m_{(1/2)^-})
\Bigg( {- 4 q_\nu \over m_{(1/2)^-}} +\gamma_\nu \Bigg)~,
\eea
where summation
over spins of the Dirac and Rarita--Schwinger spinors is performed using
\bea
\label{edhbtt05}
\sum u(q,s) \bar{u} (q,s) \es (\rlap/{q} + m_B),\nnb \\
\sum u_\mu(q,s) \bar{u}_\nu (q,s) \es (\rlap/{q} + m_B) \Bigg( g_{\mu\nu} - {1
\over 3} \gamma_\mu \gamma_\nu - {2 q_\mu q_\nu \over 3 m_B^2} + {q_\mu
\gamma_\nu - q_\nu \gamma_\mu \over 3 m_B} \Bigg)~.
\eea
If we now take into account the fact that the contributions of the spin--1/2
states are proportional to $\gamma_\mu(\gamma_\nu)$ and $q_\mu(q_\nu)$. It
follows from Eq. (\ref{edhbtt06}) that  only the structures $\rlap/{q}
g_{\mu\nu}$ and $g_{\mu\nu}$ contain contributions solely coming from
the spin--3/2 baryons, which we shall consider in
further discussion. As a result, for the physical
part of the correlator containing contributions only of the positive and negative parity
heavy spin--3/2 baryons, we get
\bea
\label{edhbtt07}
\Pi_{\mu\nu}(q) \es {\lambda^2_{(3/2)^+} \over m_{(3/2)^+}^2 - q^2} (\rlap/{q}
+ m_{(3/2)^+}) g_{\mu\nu} +
{\lambda^2_{(3/2)^-} \over m_{(3/2)^-}^2 - q^2} (\rlap/{q} -
m_{(3/2)^-})  g_{\mu\nu} + \cdots
\eea

On the QCD side, the correlation function in Eq.
(\ref{edhbtt01}) is calculated
in terms of the quark and gluon degrees of freedom  using the operator
product expansion in deep  Euclidean region, where the large and
short distance effects are separated.
After some simple calculation for the correlation function, we
obtain
\bea
\label{edhbtt08}
\Pi_{\mu\nu} (q) \es {1\over 3}
\epsilon^{abc} \epsilon^{a^\prime b^\prime c^\prime} \int d^4x
e^{iqx} \langle0| \Big\{ 4S_Q^{c b^\prime} \gamma_\nu
\widetilde{S}_{Q^\prime}^{b a^\prime} \gamma_\mu S_Q^{a c^\prime}
+2 S_Q^{c a^\prime} \gamma_\nu \widetilde{S}_{Q}^{a b^\prime}
\gamma_\mu S_{Q^\prime}^{b c^\prime} - 2S_{Q}^{c b^\prime}
\gamma_\nu \widetilde{S}_{Q}^{a a^\prime} \gamma_\mu
S_{Q^\prime}^{b c^\prime} \nnb \\
\ar 2 S_{Q^\prime}^{c a^\prime} \gamma_\nu \widetilde{S}_{Q}^{a b^\prime}
\gamma_\mu S_{Q}^{b c^\prime} - 2 S_{Q^\prime}^{c a^\prime} \gamma_\nu
\widetilde{S}_{Q}^{b b^\prime} \gamma_\mu S_Q^{a c^\prime} -
S_{Q^\prime}^{c c^\prime}  \mbox{Tr}\Big[S_{Q}^{b a^\prime} \gamma_\nu
\widetilde{S}_{Q}^{a b^\prime} \gamma_\mu \Big]\nnb\\
\ar
S_{{Q^\prime}}^{c c^\prime}  \mbox{Tr}\Big[S_{Q}^{b b^\prime} \gamma_\nu
\widetilde{S}_{Q}^{a a^\prime} \gamma_\mu \Big] -
4 S_{Q}^{c c^\prime}  \mbox{Tr}\Big[S_{Q^\prime}^{b a^\prime} \gamma_\nu
\widetilde{S}_{Q}^{a b^\prime} \gamma_\mu \Big] \Big\}|0\rangle~,
\eea
where $S_Q$ is the heavy quark operator; and $\widetilde{S} = C S^T C$.
The expression for the heavy quark propagator in $x$-representation
is given by
\bea
\label{eh32v19}
S_Q(x) &=& {m_Q^2 \over 4 \pi^2} {K_1(m_Q\sqrt{-x^2}) \over \sqrt{-x^2}} -
i {m_Q^2 \rlap/{x} \over 4 \pi^2 x^2} K_2(m_Q\sqrt{-x^2})\nnb \\& -&
ig_s \int {d^4k \over (2\pi)^4} e^{-ikx} \int_0^1
du \Bigg[ {\rlap/k+m_Q \over 2 (m_Q^2-k^2)^2} G^{\mu\nu} (ux)
\sigma_{\mu\nu} +
{u \over m_Q^2-k^2} x_\mu G^{\mu\nu} \gamma_\nu \Bigg]+\cdots~,\nnb \\
\eea
with $K_1$ and $K_2$ being the modified Bessel functions of
the second kind. The invariant functions of the structures
$\rlap/{q} g_{\mu\nu}$ or $g_{\mu\nu}$ in QCD side can be written
in terms of the dispersion relations $\Pi_i$  as
\bea
\label{edhbtt10}
\Pi_i (q) = \int ds {\rho_i(s) \over s - q^2}~,
\eea
where $i=1(2)$
corresponds to the structure $\rlap/{q} g_{\mu\nu}$ ($g_{\mu\nu}$),
and the  spectral density $\rho_i $ is given by the imaginary part
of the invariant function as
\bea
\label{nolabel04}
\rho_i(s) = {1\over \pi}
\mbox{Im}\Pi_i(s)~. \nnb
\eea
In order to calculate the invariant function we need to know
the spectral densities $\rho_1(s)$ and $\rho_2(s)$.
Using  Eq. (\ref{eh32v19}) in  Eq. (\ref{edhbtt08}), and after
lengthy calculations
 we get
\bea
\rho_1(s) \es \frac{1}{8 \pi^4}\int_{\psi_{min}}^{\psi_{max}}
\int_{\eta_{min}}^{\eta_{max}}d\psi d\eta\Bigg\{ \mu_{QQQ'}
\Bigg[2m^2_Q \psi-4m_Q m_{Q'}(-1+\psi+\eta)\nnb\\
\ar3 \eta\psi(-1+\psi+\eta)(\mu_{QQQ'}-s)\Bigg]\Bigg\}\nnb\\
\ar \frac{\langle g_s^2GG\rangle}{288\pi^4m_Q m_{Q'}}\int_{\psi_{min}}^{\psi_{max}}
\int_{\eta_{min}}^{\eta_{max}}d\psi d\eta
\Bigg\{-6(-3+4\eta)(-1+\psi+\eta)m^2_{Q'}\nnb\\
\ek 6(-1+\psi+\eta)m^2_{Q}(-3+4\psi)+m_Q m_{Q'}\Bigg[10-12\eta^2+\eta(2-60\psi)+\nnb\\
\ar (25-48\psi)\psi\Bigg]\Bigg\}~, \\ \nnb \\
\rho_2(s) \es \frac{1}{8 \pi^4}\int_{\psi_{min}}^{\psi_{max}}
\int_{\eta_{min}}^{\eta_{max}}d\psi d\eta\Bigg\{ \mu_{QQQ'}
\Bigg[3m^2_Q m_{Q'}+\eta m_{Q'}(-1+\psi+\eta)(\mu_{QQQ'}-2s)\nnb\\
\ar2m_Q\psi(-1+\psi+\eta)(\mu_{QQQ'}-2s)\Bigg]\Bigg\}\nnb\\
\ar \frac{\langle g_s^2GG\rangle}{288\pi^4m_Q m_{Q'}}\int_{\psi_{min}}^{\psi_{max}}
\int_{\eta_{min}}^{\eta_{max}}\frac{d\psi d\eta}{\eta\psi}
\Bigg\{2\psi m^2_Q m_{Q'}+\eta\Bigg[10 m^2_Q m_{Q'}\psi\nnb\\
\ar9m^3_Q(-1+\psi)\psi+m_Q m^2_{Q'}\Big[2+(7-24\psi)\psi\Big]
+6\psi^2(-1+\psi) m_{Q'}(-3\mu_{QQQ'}+5s) \Bigg]\nnb\\
\ar \eta^3\psi\Big[8m_{Q'}\psi(3 \mu_{QQQ'}-7s)+m_Q(-9\mu_{QQQ'}+12\mu_{QQQ'}\psi+15s-28\psi s)\Big]\nnb\\
\ek \eta^2\Bigg[6m^2_Qm_{Q'}\psi+2m_{Q'}\psi^2\Big[3 \mu_{QQQ'}(7-4\psi)+(-43+28\psi)s\Big]\nnb\\
\ar m_Q\Big[m^2_{Q'}(2+24\psi)-(-1+\psi)\psi(-9\mu_{QQQ'}+12\mu_{QQQ'}\psi+15s-28\psi s)
\Big]\Bigg]\Bigg\}~,
\eea
where
\bea
 \mu_{QQQ'} \es \frac{m_Q^2}{1-\psi-\eta}+\frac{m_Q^2}{\eta}+
\frac{m_{Q'}^2}{\psi}-s~,\nnb\\
 \eta_{min} \es \frac{1}{2}\Bigg[1-\psi-\sqrt{(1-\psi)\Big(1-\psi-
\frac{4\psi m_Q^2}{\psi s-m_{Q'}^2}\Big)}~~\Bigg]~,\nnb\\
\eta_{max} \es \frac{1}{2}\Bigg[1-\psi+\sqrt{(1-\psi)\Big(1-\psi-\frac{4\psi m_Q^2}{\psi
s-m_{Q'}^2}\Big)}~~\Bigg]~,\nnb\\
\psi_{min} \es \frac{1}{2s}\Bigg[s+m_{Q'}^2-4m_{Q}^2-
\sqrt{(s+m_{Q'}^2-4m_{Q}^2)^2-4m_{Q'}^2s}~~\Bigg]~,\nnb\\
\psi_{max} \es \frac{1}{2s}\Bigg[s+m_{Q'}^2-4m_{Q}^2+\sqrt{(s+m_{Q'}^2-4m_{Q}^2)^2-
4m_{Q'}^2s}~~\Bigg]~.
\eea
It should be noted here that, these expressions for the spectral
densities do not coincide with the results
presented in  \cite{R20,R21}.

Having calculated the correlation function for both physical and QCD
sides we now equate
the coefficients of the structures $\rlap/{q} g_{\mu\nu}$ and $g_{\mu\nu}$
from both sides and perform
Borel transformation with respect to $q^2$. The continuum subtraction is
done using the quark--hadron duality ansatz.
Finally, we get the following results for the sum rules:
\bea
\label{edhbtt13}
\lambda_{(3/2)^+}^2 e^{-m_{(3/2)^+}^2/M^2} +
\lambda_{(3/2)^-}^2 e^{-m_{(3/2)^-}^2/M^2} \es
\int_{(2m_Q+m_Q^\prime)^2}^{s_0} ds \rho_1(s)e^{-s/M^2}~,\nnb \\ \\
\label{edhbtt14}
\lambda_{(3/2)^+}^2 m_{(3/2)^+} e^{-m_{(3/2)^+}^2/M^2} -
\lambda_{(3/2)^-}^2 m_{(3/2)^-} e^{-m_{(3/2)^-}^2/M^2} \es
\int_{(2m_Q+m_Q^\prime)^2}^{s_0} ds \rho_2(s)
e^{-s/M^2}~,\nnb\\
\eea
where $M^2$ and  $s_0$ are Borel mass parameter and continuum threshold,
respectively.
These equations contain four unknowns:
$ \lambda_{(3/2)^+}$, $m_{(3/2)^+}$,  $ \lambda_{(3/2)^-}$ and
$m_{(3/2)^-}$.
Hence we need two more equations in order to solve for these quantities.
Two more equations can be found by taking derivatives of
both sides of the above equations with respect to $-1/M^2$, which
gives:
\bea
\label{edhbtt133}
\lambda_{(3/2)^+}^2 m^2_{(3/2)^+}e^{-m_{(3/2)^+}^2/M^2} +
\lambda_{(3/2)^-}^2 m^2_{(3/2)^-}e^{-m_{(3/2)^-}^2/M^2} \es
\int_{(2m_Q+m_Q^\prime)^2}^{s_0} ds~s \rho_1(s)e^{-s/M^2}~,\nnb \\ \\
\label{edhbtt144}
\lambda_{(3/2)^+}^2 m^3_{(3/2)^+} e^{-m_{(3/2)^+}^2/M^2} -
\lambda_{(3/2)^-}^2 m^3_{(3/2)^-} e^{-m_{(3/2)^-}^2/M^2} \es
\int_{(2m_Q+m_Q^\prime)^2}^{s_0} ds~s \rho_2(s)
e^{-s/M^2}~.\nnb\\
\eea
Solving equations (\ref{edhbtt13}), (\ref{edhbtt14}),
(\ref{edhbtt133}) and (\ref{edhbtt144}) simultaneously, one can
find the four unknowns
$\lambda_{(3/2)^+}$, $m_{(3/2)^+}$,  $ \lambda_{(3/2)^-}$ and $m_{(3/2)^-}$.

At the end of this section we would like to make the following remark about
the radiative ${\cal O}(\alpha_s) $ corrections to the spectral densities. These corrections modify the perturbative parts  by the factor of
$1+\frac{\alpha_s}{\pi} f(m_Q,m_Q',s)$, where $f(m_Q,m_Q',s)$ is a function of quark masses and $s$. The mass of baryons from sum rules is
determined by the ratio of the two corresponding spectral densities. Therefore, even if the radiative corrections are large, 
they  can not change the values of the masses, considerably. Because these two large corrections are
practically cancel each other. Formally, these corrections can be absorbed by the pole residues.  

\section{Numerical results}

In performing the numerical analysis of the sum rules for the masses and
residues of the triply heavy spin--3/2  baryons, we need the values
of the input parameters entering into the sum rules.
For the heavy quark masses we use their pole values,
$m_b=(4.8 \pm 0.1)~GeV$ and $m_c=(1.46 \pm 0.05)~GeV$
\cite{R23}. The numerical value of the gluon
condensate is taken to be $\langle g_s^2GG \rangle=4 \pi^2 (0.012\pm
0.004)~GeV^4$ \cite{R23}. It should be noted here that, 
if instead of the pole mass values of the heavy quarks 
their $\overline{MS}$ \cite{R24} values are used,
our analysis shows that the results for the masses do not change considerably. 

The sum rules for the masses and residues  contain two
auxiliary parameters, namely continuum threshold $s_0$ and Borel
mass parameter $M^2$. Obviously, the physical quantities should be
independent of the variations of these auxiliary parameters.
The continuum threshold  is not totally arbitrary but it is
correlated with  the energy of the first excited state in each
channel. It is usually chosen as $\sqrt{s_0} = (m_{ground}+0.5)~GeV$,
and the domain of those values of $s_0$ is searched which reproduces
this relation. As the result of using this requirement we have obtained the
intervals of $s_0$ for each baryon, and presented them in Table 2.
Numerical analysis shows that our results on the masses are weakly
dependent on the variations in $s_0$ in the considered interval.

The working region for the Borel mass parameter $M^2$ is found as
follows. The upper bound on this parameter is obtained by
requiring  that the pole contribution to the sum rules
exceeds   the contributions of the higher states and continuum,
i.e., the condition,
\bea
\label{nolabel}
{\ds \int_{s_0}^{\infty}\ds  ds \rho(s)
e^{-s/M^2} \over \ds \int_{s_{min}}^\infty ds \rho(s) e^{-s/M^2}}
< 1/3~,
\eea
should be satisfied. The lower bound on $M^2$ is
obtained by demanding that the contribution of the perturbative
part exceeds   the non-perturbative contributions. From these
restrictions we obtain the working regions for the Borel mass
parameter for all members of the triply heavy baryons, which are also presented
in Table 2.

\begin{table}[h]

\renewcommand{\arraystretch}{1.3}
\addtolength{\arraycolsep}{-0.5pt} \small
$$
\begin{array}{|l|r@{\,-\,}l|r@{\pm}l|r@{\pm}l|r@{\pm}l|}
\hline \hline
%\mbox{\small{~~~\,transition}}
&    \multicolumn{2}{c|}{\mbox{$M^2(GeV^2)$}}  &
\multicolumn{2}{c|}{\mbox{$\sqrt{s_0}(GeV)$}} &
\multicolumn{2}{c|}{\mbox{$m(GeV)$}} &
\multicolumn{2}{c|}{\mbox{$\lambda(GeV^3)$}}       \\ \hline
\Omega_{ccc}({{3 \over 2}^+})             &  4.5&8.0   &  5.6&0.2   &   4.72&0.12   & 0.09&0.01 \\
\Omega_{ccc}({{3 \over 2}^-})             &  4.5&8.0   &  5.8&0.2   &   4.9&0.1     & 0.11&0.01 \\
\Omega_{ccb}({{3 \over 2}^+})             &  6.0&10.0  &  8.8&0.2   &   8.07&0.10   & 0.06&0.01 \\
\Omega_{ccb}({{3 \over 2}^-})             &  6.0&10.0  &  9.0&0.2   &   8.35&0.10   & 0.07&0.01 \\
\Omega_{bbc}({{3 \over 2}^+})             &  8.0&10.5  & 12.0&0.2   &  11.35&0.15   & 0.08&0.01 \\
\Omega_{bbc}({{3 \over 2}^-})             &  8.0&10.5  & 12.2&0.2   &  11.5&0.2     & 0.09&0.01 \\
\Omega_{bbb}({{3 \over 2}^+})             & 12.0&18.0  & 15.3&0.2   &  14.3&0.2     & 0.14&0.02 \\
\Omega_{bbb}({{3 \over 2}^-})             & 12.0&18.0  & 15.5&0.2   &  14.9&0.2     & 0.20&0.02 \\
 \hline \hline
\end{array}
$$
\caption{ Working regions of auxiliary parameters  $M^2$ and $s_0$
together  with the masses $m$ and residues $\lambda$ of the triply
heavy spin--3/2 baryons. In the numerical analysis we use pole mass
of the heavy quarks.}
\renewcommand{\arraystretch}{1}
\addtolength{\arraycolsep}{-1.0pt}

\end{table}

Having determined the working regions for Borel mass parameter
$M^2$ entering the sum rules, now we are ready to calculate of the masses and
residues of the corresponding triply heavy baryons. As example, in Figs. (1)
and (2) we present the dependence of the mass of
$\Omega_{ccc}({3^+\over 2})$ and $\Omega_{ccc}({3^-\over 2})$ baryons on the
Borel mass parameter $M^2$. We deduce from these figures that the masses of
the $\Omega_{ccc}({3^+\over 2})$ and $\Omega_{ccc}({3^-\over 2})$ baryons
are equal to $(4.72 \pm 0.12)~GeV$ and $(4.9 \pm 0.1)~GeV$, respectively. We
have performed similar analysis for the other triply heavy baryons. The
results for the masses and residues of all members of the triply heavy.
spin-3/2 baryons of both parities are presented in Table 2.
From this Table we see that, for  the $\Omega$ baryons
the masses of the negative parity baryons are
slightly greater than those of the positive parity baryons.
In the case of residues, also,
the negative parity baryons have residues slightly higher than
those of positive parity baryons.

Now, we compare our results on the masses and residues obtained
from using the pole masses of the quarks with the existing
predictions of other theoretical approaches. First, in Table 3, we
compare our results on the masses  with existing predictions of
approaches like lattice calculation, QCD bag model, variational
method, modified bag model, relativistic quark model, non-relativistic
quark mode, QCD sum rules and lattice calculations
\cite{R08,R09,R10,R11,R12,R13,R14,R15,R20,R21}.
In all cases the predictions for the masses of the $ccc$, $ccb$ and
$bbc$ baryons are, roughly, in good agreement within the
error limits, except than the results of \cite{R20}, which are
considerably low compared to the other predictions. For the $bbb$
baryons, our results are approximately consistent with the
existing results of \cite{R10,R12,R13}, however, our
predictions on these baryons are lower compared to the
existing predictions of \cite{R11,R14,R15,R21}. In
the meanwhile, our result on the positive parity $bbb$ baryon is
considerably high compared to the result of \cite{R20}. The  differences
between our results for the masses and those
presented in  \cite{R20} can be due to the following  reasons. Firstly,
in \cite{R20} the
contributions of the negative parity baryons are not taken into
account. Secondly, our spectral densities are different than those
presented
in \cite{R20} for positive parity baryons. Note that  predictions
for the residues of triply heavy baryons are absent at all in
\cite{R20}.

\begin{table}[h]

\renewcommand{\arraystretch}{1.3}
\addtolength{\arraycolsep}{-0.5pt} \small
$$
\begin{array}{|l|r@{\pm}l|r@{\pm}l|c|c|c|c|c|c|c|c|c|}
\hline \hline
%\mbox{\small{~~~\,transition}}
&    \multicolumn{2}{c|}{\mbox{Our work}}  &
\multicolumn{2}{c|}{\mbox{\cite{R21}}}  &
\multicolumn{1}{c|}{\mbox{\cite{R10}}}        &
\multicolumn{1}{c|}{\mbox{\cite{R11}}}        &
\multicolumn{1}{c|}{\mbox{\cite{R12}}}         &
\multicolumn{1}{c|}{\mbox{\cite{R13}}}        &
\multicolumn{1}{c|}{\mbox{\cite{R14}}}        &
\multicolumn{1}{c|}{\mbox{\cite{R20}}}        &
\multicolumn{1}{c|}{\mbox{\cite{R15}}}        &
\multicolumn{1}{c|}{\mbox{\cite{R09}}}      &
\multicolumn{1}{c|}{\mbox{\cite{R08}}}      \\ \hline
m_{\Omega_{ccc}({{3 \over 2}^+})}             &  4.72&0.12  &  4.99&0.14   & 4.79  & 4.925  &  4.76 &  4.777 &  4.803 &  4.67\pm 0.15 &  4.965 & 4.7   &       \\
m_{\Omega_{ccc}({{3 \over 2}^-})}             &  4.9&0.1    &  5.11&0.15   &       &        &       &        &        &               &        & 5.1   &       \\
m_{\Omega_{ccb}({{3 \over 2}^+})}             &  8.07&0.10  &  8.23&0.13   & 8.03  & 8.200  &  7.98 &  8.005 &  8.025 &  7.45\pm 0.16 &  8.265 & 8.05  &       \\
m_{\Omega_{ccb}({{3 \over 2}^-})}             &  8.35&0.10  &  8.36&0.13   &       &        &       &        &        &               &        &       &       \\
m_{\Omega_{bbc}({{3 \over 2}^+})}             & 11.35&0.15  & 11.49&0.11   & 11.20 & 11.480 & 11.19 & 11.163 & 11.287 & 10.54\pm 0.11 & 11.554 &       &       \\
m_{\Omega_{bbc}({{3 \over 2}^-})}             & 11.5&0.2    & 11.62&0.11   &       &        &       &        &        &               &        &       &       \\
m_{\Omega_{bbb}({{3 \over 2}^+})}             & 14.3&0.2    & 14.83&0.10   & 14.30 & 14.760 & 14.37 & 14.276 & 14.569 & 13.28\pm 0.10 & 14.834 &       & 14.37 \\
m_{\Omega_{bbb}({{3 \over 2}^-})}             & 14.9&0.2    & 14.95&0.11   &       &        &       &        &        &               &        &       & 14.72 \\
\hline \hline
\end{array}
$$
\caption{The masses of the triply heavy spin--3/2 baryons in units
of $GeV$, calculated using the pole masses of the heavy quarks,
compared with other theoretical predictions.}
\renewcommand{\arraystretch}{1}
\addtolength{\arraycolsep}{-1.0pt}

\end{table}

The comparison of our results on the residues of the triply heavy
baryons with those of the \cite{R21} as the only existing
results in the literature is made in Table 4. From this Table we
observe that the predictions of \cite{R21} on the
residues are approximately (2-5)
times grater than our results depending on the quark contents of
the baryons. These differences can be attributed to the different
interpolating currents that have been used in these two works. 
Moreover, in determination
of the masses and residues of the negative and
positive parity baryons we have coupled four equations but in
\cite{R21}, only two
equations  exist. In the case of masses,  the ratios of two corresponding
sum rules are considered and therefore the errors cancel each other.
For this
reason our predictions for the masses are comparable with  those of
\cite{R21}. But in
determination of the residues,  there are no ratios of sum rules and one
of the obtained
equations are used, which leads to the above-mentioned considerable
differences. Although triply heavy baryons have not yet been discovered in
experiments, their production at LHCb has theoretically been studied in \cite{R25}, and it
is found that $10^4$-$10^5$ events of triply heavy baryons can be produced at
$10~fb^{-1}$ integrated luminosity.

\begin{table}[h]

\renewcommand{\arraystretch}{1.3}
\addtolength{\arraycolsep}{-0.5pt} \small
$$
\begin{array}{|l|r@{\pm}l|r@{\pm}l|}
\hline \hline
%\mbox{\small{~~~\,transition}}
&    \multicolumn{2}{c|}{\mbox{Present study}}  &
\multicolumn{2}{c|}{\mbox{\cite{R21}}}      \\ \hline
\lambda_{\Omega_{ccc}({{3 \over 2}^+})}             &  0.09&0.01  &   0.20&0.04    \\
\lambda_{\Omega_{ccc}({{3 \over 2}^-})}             & 0.11&0.01 &   0.24&0.04          \\
\lambda_{\Omega_{ccb}({{3 \over 2}^+})}             &  0.06&0.01  &   0.26&0.05     \\
\lambda_{\Omega_{ccb}({{3 \over 2}^-})}             &   0.07&0.01  &   0.32&0.06           \\
\lambda_{\Omega_{bbc}({{3 \over 2}^+})}             &  0.08&0.01  &  0.39&0.09     \\
\lambda_{\Omega_{bbc}({{3 \over 2}^-})}             &  0.09&0.01  &  0.49&0.10           \\
\lambda_{\Omega_{bbb}({{3 \over 2}^+})}             & 0.14&0.02  &  0.68&0.16    \\
\lambda_{\Omega_{bbb}({{3 \over 2}^-})}             &  0.20&0.02  &  0.86&0.17          \\
\hline \hline
\end{array}
$$
\caption{The residues of the triply heavy spin--3/2 baryons in
units of $GeV$, calculated using the pole masses of the heavy
quarks, compared with the predictions of \cite{R21}.}
\renewcommand{\arraystretch}{1}
\addtolength{\arraycolsep}{-1.0pt}

\end{table}

In summary, we evaluated the masses and residues of the triply
heavy spin--3/2 baryons   with both
positive and negative parities in the framework of the QCD sum
rules as one of the most powerful non-perturbative method.
The results obtained in this work are compared with
the predictions of other theoretical approaches.

We hope it would be possible
to measure the masses and decays of the triply heavy baryons in
the near future at LHCb.

\begin{figure}[h]
\centering
\begin{tabular}{ccc}
\includegraphics[totalheight=18cm,width=15cm]{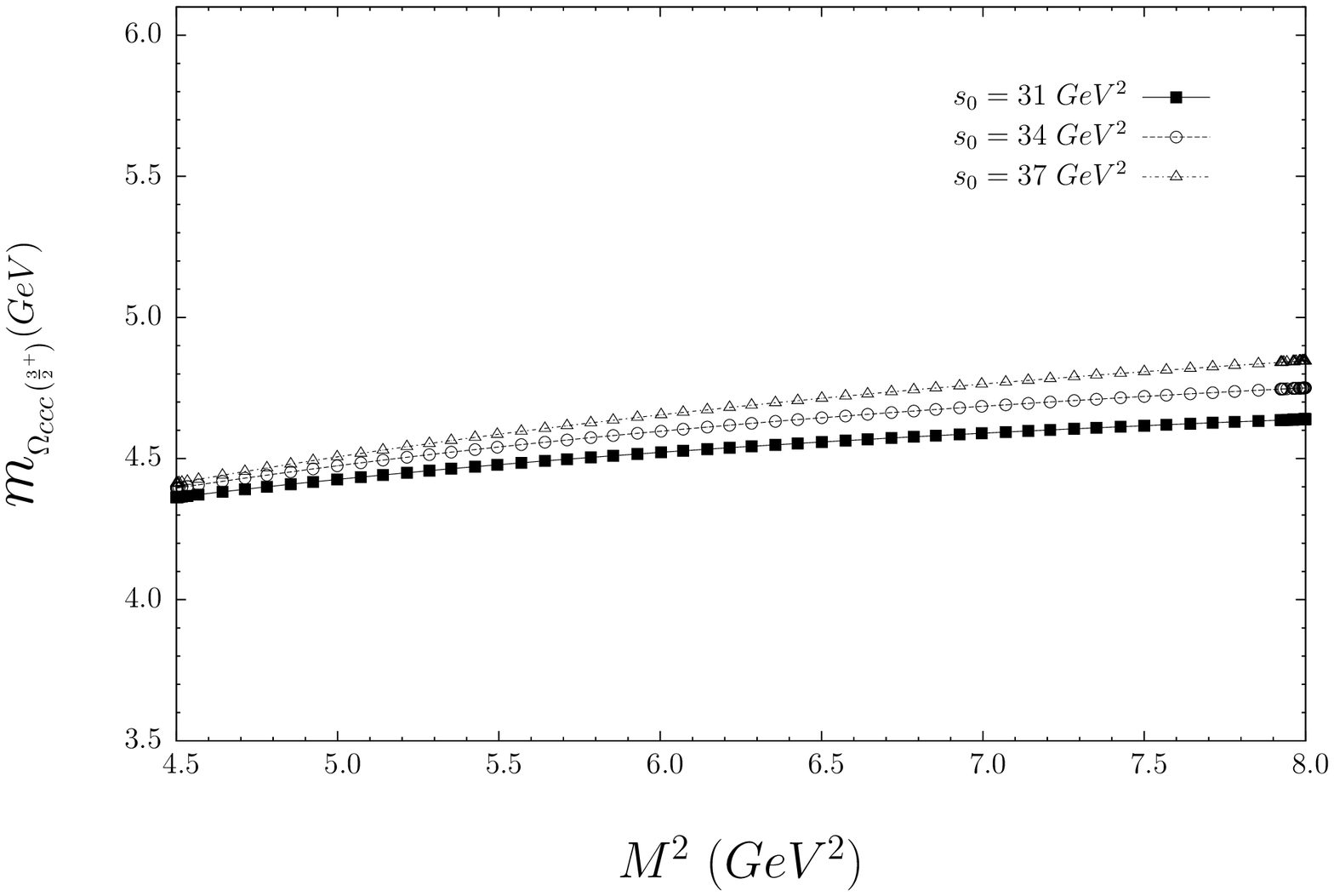}
\end{tabular}
\caption{Dependence of the mass of the triply heavy, positive parity
${\Omega_{ccc}({{3 \over 2}^+})}$ baryon on the auxiliary
Borel mass parameter $M^2$, at several fixed values of the
continuum threshold $s_0$.}
\end{figure} 

\begin{figure}[h]
\centering
\begin{tabular}{ccc}
\includegraphics[totalheight=18cm,width=15cm]{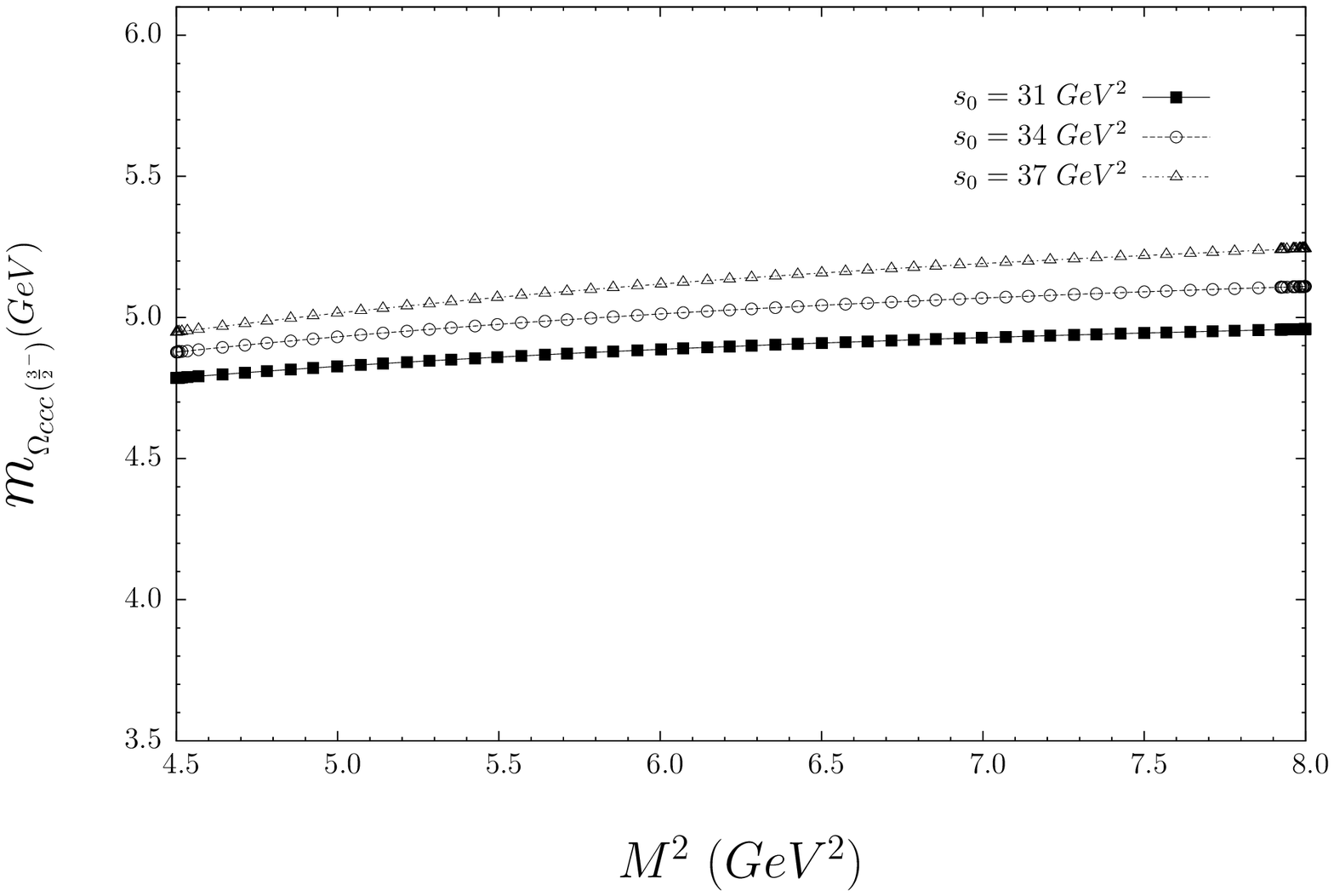}
\end{tabular}
\caption{The same as Fig. 1, but for the negative parity
${\Omega_{ccc}({{3 \over 2}^-})}$ baryon.}
\end{figure} 
%\begin{figure}
%\vskip 3. cm
 %   \special{psfile=fig1.eps hoffset= -27 voffset=-540 hscale=80 vscale=80
%angle=0}
%\vskip 7.0cm
%\caption{Dependence of the mass of the triply heavy, positive parity
%${\Omega_{ccc}({{3 \over 2}^+})}$ baryon on the auxiliary
%Borel mass parameter $M^2$, at several fixed values of the
%continuum threshold $s_0$.}
%\begin{center}  
%{\bf Fig. 1-a}   
%\end{center}   
%\end{figure}

%\begin{figure}
%\vskip 3. cm
%    \special{psfile=fig2.eps hoffset= -27 voffset=-540 hscale=80 vscale=80
%angle=0}
%\vskip 7.0cm
%\caption{The same as Fig. 1, but for the negative parity
%${\Omega_{ccc}({{3 \over 2}^-})}$ baryon.}
%\begin{center}
%{\bf Fig. 1-a}   
%\end{center}
%\end{figure}


\newpage


\begin{thebibliography}{99}

\bibitem{R01} T. Kuhr,  arXiv:1109.1944 [hep-ex].

\bibitem{R02} S. Chatrchyan {\it et. al}, CMS Collaboration,
  Phys. Rev. Lett.  108, 252002 (2012).

\bibitem{R03} M. Mattson {\it et. al}, SELEX Collaboration,
  Phys. Rev. Lett. 89, 112001 (2002).

\bibitem{R04} A. Ocherashvili {\it et. al},
  Phys. Lett. B  628, 18 (2005).

\bibitem{R05} J. Eigelfried {\it et. al}, SELEX Collaboration,
  Nucl. Phys. A  752, 121 (2005).

\bibitem{R06} N. Brambilla, T. Roesch, A. Vairo,
  Phys. Rev.  D 72, 034021 (2005).

\bibitem{R07} T. W. Chiu, T. H. Hsieh,
  Nucl. Phys.  A 755, 471 (2005).

\bibitem{R08} S. Meinel,
  Phys. Rev.  D 85, 114510 (2012).

\bibitem{R09} M. Padmanath, R. G. Edwards, N. Mathur,
  arXiv:1307.7022 [hep-lat];
  arXiv:1311.4354 [hep-lat].

\bibitem{R10}  P. Hasenfratz, R. R. Horgan, J. Kuti, J. M. Richard,
  Phys. Lett.  B 94, 401 (1980).

\bibitem{R11}  J. D. Bjorken,
  Preprint FERMILAB-Conf-85-069, C85-04-20 (1985).

\bibitem{R12} Y. Jia,
  JHEP, 10, 073  (2006).

\bibitem{R13} A. Bernotas and V. Simonis,
  Lith. J. Phys.  49, 19 (2009).

\bibitem{R14} A. P. Martynenko,
  Phys. Lett.  B 663, 317 (2008).

\bibitem{R15} W. Roberts and M. Pervin,
  Int. J. Mod. Phys. A 23 , 2817 (2008).

\bibitem{R16}  J. Vijande, H. Garcilazo, A. Valcarce and F. Fernandez,
  Phys. Rev. D 70, 054022 (2004).

\bibitem{R17} B. Patel, A. Majethiya, P. C. Vinodkumar,
  Pramana,  72, 679  (2009).

\bibitem{R18} F. J. Llanes-Estrada, O. I. Pavlova, R. Williams,
  Eur. Phys. J. C 72, 2019  (2012).

\bibitem{R19} X. H. Guo, K. W. Wei, X. H. Wu,
  Phys. Rev. D 78, 056005 (2008).

\bibitem{R20} J. R. Zhang and M. Q. Huang,
  Phys. Lett. B 674, 28 (2009).

\bibitem{R21} Zhi-Gang Wang,
  Commun. Theor. Phys. 58, 723 (2012).

\bibitem{R22} T. M. Aliev, K. Azizi, M. Savci, JHEP 1304, 042 (2013).

\bibitem{R23} P. Colangelo, A. Khodjamirian,
  ``At the Frontier of Particle Physics/Handbook of QCD'',
  edited by M. Shifman (World Scientific, Singapore, 2001),
  Vol. 3, p. 1495.

\bibitem{R24} A. Khodjamirian, Ch. Klein, Th. Mannel, and N. Offen,
  Phys. Rev. D 80, 114005 (2009).

\bibitem{R25} Yu-Qi Chen, Su-Zhi Wu,
  JHEP 1108 (2011) 144, Erratum-ibid. 1109 (2011) 089.


\end{thebibliography}
\end{document}